\documentclass[fleqn,usenatbib]{mnras}

\usepackage{newtxtext,newtxmath}

\usepackage[T1]{fontenc}

\DeclareRobustCommand{\VAN}[3]{#2}
\let\VANthebibliography\thebibliography
\def\thebibliography{\DeclareRobustCommand{\VAN}[3]{##3}\VANthebibliography}

\usepackage{graphicx}	
\usepackage{amsmath}

\title[H-band spectral fluctuations]{Decomposing cool stellar populations 
with H-band spectral fluctuations: Long-period variable stars in NGC\,5128 and carbon stars in NGC\,5102}

\author[R.J. Smith \& A. Gvozdenko]{
Russell J. Smith\thanks{E-mail: russell.smith@durham.ac.uk}  and
Anastasia Gvozdenko
\\
Centre for Extragalactic Astronomy, Department of Physics, Durham University, South Road, Durham DH1 3LE, UK\\
}

\date{Accepted 21 May 2026. Received 20 May 2026; in original form 2 April 2026.}

\pubyear{\the\year{}}

\defcitealias{2022MNRAS.509.5737S}{S22}
\newcommand{\stt}{\citetalias{2022MNRAS.509.5737S}}

\begin{document}
\label{firstpage}
\pagerange{\pageref{firstpage}--\pageref{lastpage}}
\maketitle

\begin{abstract}
We analyse new H-band integral-field unit observations of two galaxies at $\sim$4\,Mpc, using a principal components analysis of pixel spectra to probe their giant star content.
 In both galaxies, the signals arise in near-resolved point-like sources without large-scale variation, consistent with each pixel sampling stars randomly from a common underlying population.
In the (mostly) old bulge of NGC\,5128, the observed pixel-to-pixel variation is dominated by a component with a mid-M giant spectrum with prominent CO bandheads. We also  recover a smoother second spectral component, apparently driven by contributions from later spectral types. This component is not present in predictions from Poisson-sampled models of old stellar populations; we suggest that it arises from the cool phases of long-period variable stars. (An appendix provides direct evidence for such variables in complementary two-epoch MUSE observations.) In the contrasting galaxy NGC\,5102, where a post-starburst stellar population is known to be present, we again find two distinct components. As before, the first component carries the CO bands typical of M-giants.  The second eigenspectrum in this younger galaxy shows a strong 1.77\,$\mu$m C$_2$ bandhead, a feature which is characteristic of carbon stars. Our results highlight the ability of integral field data to access information beyond the total spectrum, even when individual stars cannot be resolved.
\end{abstract}


\section{Introduction}

The study of resolved stellar populations, whether by photometry and colour-magnitude diagrams or through spectroscopy, is a mainstay of observational astrophysics. 
A new generation of instruments and telescopes will extend the limit of resolved population work to distances beyond a few Mpc, with the {\it James Webb Space Telescope} (JWST) already yielding photometry of 
giant-branch stars in Virgo Cluster ellipticals,  \citep{2025ApJ...982...26A}, while the 30m-class telescopes will provide resolved spectroscopy \citep[][]{2018arXiv181004422G}.
Notwithstanding such advances, however, some important astrophysical environments will remain unresolvable for the foreseeable future, and at any given time, other applications will be beyond the ambit of traditional resolved study. In such cases, we are usually limited to integrated-light methods, relying heavily on synthesis models to infer the relative contributions of different types of star \citep{2013ARA&A..51..393C,2020Galax...8....6S}.
Inevitably, such models are subject to uncertainty, ambiguity and incompleteness, especially for evolutionary phases beyond the helium flash. 

The central regions of giant ellipticals provide one example of stellar populations that will not be truly resolvable by any currently imaginable facility, but which are of great interest given their unusual properties. Today's ellipticals harbour stars formed at the earliest epochs and under conditions that may differ markedly from any environment observable at close quarters. JWST observations are revealing unusual gas-phase metallicities in the likely progenitors of the oldest galaxies \citep[][]{2023MNRAS.523.3516C}, while at the present day, ellipticals are characterised by high enhancements in $\alpha$-elements, and in sodium, and spectral signatures apparently requiring a bottom-heavy initial mass function \citep[][]{Conroy2012,2013MNRAS.429L..15F,2017MNRAS.464.3597L}. 
The infra-red spectral features in old ellipticals, including the very strong CO bandheads, have proven difficult to match consistently with the optical lines \citep{Elham2022,2025A&A...700A..64L} in current models. 
Further peculiarities in ellipticals are hinted at by the ultraviolet excess flux, caused apparently by extreme horizontal branch stars \citep{1999ARA&A..37..603O}. At high metallicity, such stars are easier to produce if their progenitors are enriched in helium \citep*{2017ApJ...842...91C}, which itself can be generated by the same high-temperature proton-capture processes implicated in sodium enhancements, e.g. in globular cluster populations \citep{2002A&A...395...69D}. This raises the intriguing possibility that some fraction of stars in ellipticals may be anomalously helium rich \citep{2011ApJ...740L..45C,2018ApJ...857...16G,2024ApJ...966...50A}. 
If so, then they would evolve faster at fixed mass, and hence lower-mass stars would be passing through any evolutionary phase, compared to standard models. 
The characteristics of such populations are 
not testable in any other high metallicity environment, making ellipticals a unique test of unusual stellar physics. 

The  problems above relate mainly to classic ancient ellipticals, but younger early-type galaxies also pose challenges and opportunities for testing stellar population models.  In galaxies with intermediate-age populations, the impact of thermally-pulsing asymptotic giant branch (TP-AGB) stars has long been debated \citep[e.g.][]{Baldwin2018,DahmerHahn2018}, although recent observations of distant galaxies are limiting the viable recipes \citep{2025NatAs...9..128L}. Newly-quenched ``post-starburst'' galaxies in the local Universe offer a different route to testing models for these stars.

Although the centres of ellipticals are not resolvable, they are nonetheless amenable to `fluctuation' methods, beyond standard integrated light spectroscopy, which can extract information at least about the most luminous stars present. These approaches extract information from  pixel-to-pixel differences in the random sampling of stars from the underlying population. In the imaging domain, this is the method of surface brightness fluctuations (SBF), which measures a moment of the stellar luminosity function heavily weighted to the brightest giants \citep{1988AJ.....96..807T}. Beyond its application as a distance estimator, the SBF method is sensitive as a probe of stellar populations \citep[e.g.][]{2005AJ....130.2625R}.
Several authors have considered extensions of the fluctuation approach into the spectroscopic regime. For example, \cite{2014ApJ...797...56V} computed the spectral ratio between `brighter' and `fainter' pixels in a Poisson-sampled population, and showed that narrow-band imaging could access some of the information carried. Alternatively, \cite{2018MNRAS.480..629M} applied the imaging SBF method to each wavelength channel of an integral-field unit (IFU) observation, extracting true spectral-domain information beyond the ordinary integrated-light spectrum.

\citet[][hereafter \stt]{2022MNRAS.509.5737S} introduced a more general approach, 
in which many spectra from an IFU datacube, considered without reference to their spatial configuration, are subjected to a principal components analysis (PCA) to identify the most informative variation signatures. Through application to stellar population models, it was shown that the leading eigenspectra could be related fairly straightforwardly to the underlying stellar spectra of the brightest contributing stars. \stt\ demonstrated the practical feasibility of this technique using  high-angular-resolution optical observations of the nearby peculiar early-type galaxy NGC\,5128, acquired with MUSE\footnote{The Multi-Unit Spectroscopic Explorer  \citep{2010SPIE.7735E..08B}, at the 8.2-m European Southern Observatory Very Large Telescope (VLT).}. Given the structure of the target, it was necessary to observe a field at some distance from the centre, but the population was still essentially unresolved. From the MUSE data,  it was possible to recover only one meaningful spectral component, with an eigenspectrum  showing the expected features (mainly Ti\,O bands) 
of cool M-giants. 

In this paper, we present an extension of the \stt\ approach, applied to H-band spectroscopy obtained with ERIS\footnote{The Enhanced Resolution Imager and Spectrograph \citep{2023A&A...674A.207D}, also at the VLT.}. In related work (Gvozdenko \& Smith, {\it in preparation}, hereafter GS26) we have also extended the range of model predictions over the whole optical/near-infrared spectral region and for a broader range of stellar populations.
As well as revisiting the (dominantly) old  bulge population of  NGC\,5128, we present results for the contrasting case of its companion galaxy NGC\,5102, which is known to harbour a post-starburst population, with signatures of carbon-rich TP-AGB stars in its near-infrared spectrum \citep{2011ApJ...727L..15M,2015ApJ...799...97D}.

In the remainder of the paper, we describe the acquisition and processing of the observational data in Section~\ref{sec:data} and the PCA  analysis and results in Section~\ref{sec:pca}. We discuss the interpretation of our results Section~\ref{sec:disc} and draw brief conclusions in Section~\ref{sec:concs}.

\section{Observations and data reduction}\label{sec:data}

\begin{figure*}
 \includegraphics[width=177mm]{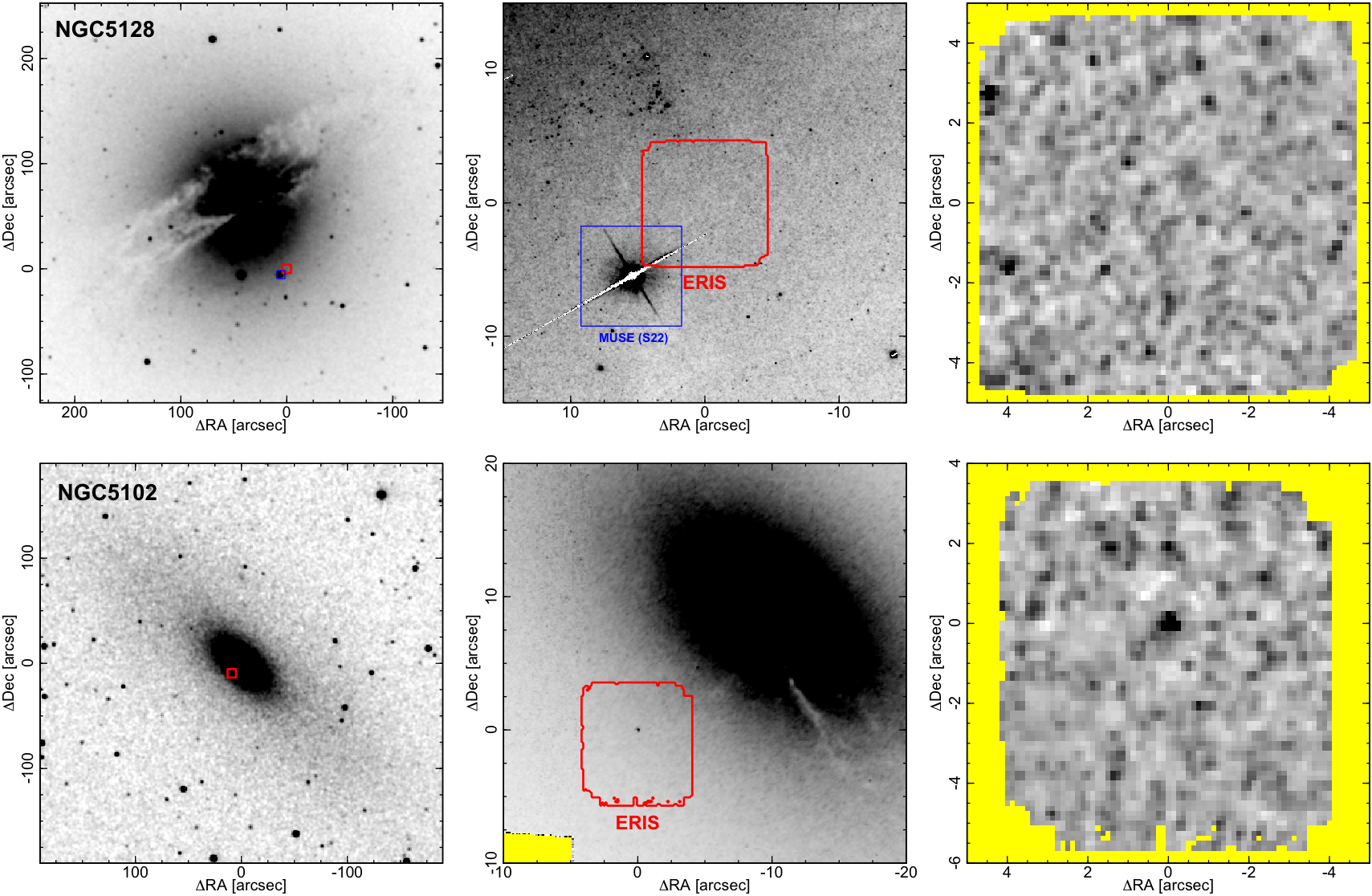}
 \vskip 0.5mm
 \caption{The fields observed with ERIS (red) for NGC\,5128 (upper panels) and NGC\,5102 (lower panels). Also marked is the MUSE-NFM field of S22 in NGC\,5128 (blue).
The first panel of each row shows a wide field J-band image from \protect\cite{2006AJ....131.1163S}, while the second is an archival optical HST image: ACS/F606W for NGC\,5128 \citep{2006AJ....132.2187H}; WFPC2/F569W for NGC\,5102 \citep{Mitzkus2017}. In the third panel, we show the collapsed ERIS image, including only pixels with at least three exposures (900\,s).
 }
 \label{fig:fields}
\end{figure*}

We observed one field in NGC\,5128, and one in NGC\,5102, using the ERIS integral-field unit mode, with laser guide star adaptive optics (AO). We used the configuration with the largest plate scale, providing a 8$\times$8\,arcsec$^2$ field-of-view. The data were acquired with the low-resolution grating, covering the whole H-band window at resolving power $R$\,$\sim$\,5000. The field locations are shown in Figure~\ref{fig:fields}. The spatial sampling is set by the 0.250\,arcsec slice width in one direction, and by the 0.125\,arcsec$^2$ detector pixel scale along the slit. Some of the exposures were acquired with a 90-degree position angle rotation,
to reduce anisotropic sampling to some extent.

For NGC\,5128, the field was chosen to overlap partially with the MUSE field of S22. (Full overlap was not possible to avoid persistence effects from the bright star used for the on-axis MUSE AO correction.) The field samples a representative part of the bulge stellar population, avoiding the young stars closer to the disk, and avoiding also the dust structures evident in {\it Hubble Space Telescope} (HST) imaging. The NGC\,5128 data used here comprise a total of 40 dithered on-source exposures, each of 300\,s, for a total of 12000\,s maximum integration. No dedicated blank sky frames were acquired.
For a distance of 3.8\,Mpc \citep{2010PASA...27..457H}, the H-band surface brightness of 16.7\,mag\,arcsec$^{-2}$ in this field corresponds to an average 11000 solar luminosities per 0.125$\times$0.125\,arcsec$^2$ ERIS pixel. For comparison, GS26 report a SBF luminosity $\sim$2000\,$L_\odot$ for old population models, i.e. variations are expected to be driven by only a few stars per pixel.
Moreover, the very brightest stars on the isochrone reach $\sim$10000\,$L_\odot$, so they will be present in a minority of pixels, and hence be effectively resolved.
 
For NGC\,5102, the ERIS 
field was placed at $\sim$15\,arcsec along the minor axis, constrained largely by AO reference star availability. Dedicated offset-sky frames were alternated with the science pointing, though in practice the sky frames were not used in the data reduction (see below). We obtained 
twelve 300\,s on-source exposures for a maximum total integration time of  3600\,s. 
At the observed field position, the H-band surface brightness of 17.2\,mag\,arcsec$^{-2}$ corresponds to an average of 7300 solar luminosities per 0.125$\times$0.125\,arcsec$^{2}$ ERIS pixel,  
for an adopted distance of 4.0\,Mpc \citep{2017MNRAS.464.4789M}.

Each individual on-source exposure was reduced using the standard ESO-provided ERIS reduction pipeline. For the NGC\,5128 observation, no dedicated blank sky exposures were obtained. Since it is the pixel-to-pixel {\it differences} that are of scientific interest, the `uniform' background can be absorbed into the subtracted sky spectrum. For simplicity and consistency, we applied the same data reduction scheme to the NGC\,5102 observations, despite having acquired offset sky frames in that instance.
Subtracting a background correction derived from the median of all dithered observations within a given 8- or 4-frame observing block left large, spatially-structured, residuals around the OH airglow lines. To improve on this, we implemented a custom sky-correction method, incorporating a slice-wise median background subtraction, followed by a PCA reconstruction of the sky residuals \citep[following the general approach of][]{2005MNRAS.358.1083W}, with 16 retained components. A final step flagged sporadic bad pixels and cosmic ray hits. This process was applied to each of the individual reduced exposures. 
The individual datacubes were then 
converted to square 0.125$\times$0.125\,arcsec$^2$ pixels, and aligned with simple integer-pixel shifts (determined through cross-correlation), to avoid introducing any further noise correlation, beyond that imposed by the originally rectangular pixels.
An approximate telluric absorption correction was obtained from the library of transmission functions provided by ESO, convolved and aligned to match the water absorption features which were otherwise visible in the derived eigenspectra. 

Collapsed images of the final stacked data are shown in the right-hand panels of  Figure~\ref{fig:fields}. As expected from the surface brightness argument above, the fields appear marginally resolved, with individual giants apparently identifiable in the collapsed image, though they are too faint to extract spectra for any of them individually. The NGC\,5102 field arguably appears more resolved, with fewer, brighter sources. This is qualitatively consistent with the expectation of that brighter AGB stars dominate the H-band light in younger stellar populations.

The NGC\,5102 field was selected to include a prominent star from which the point-spread function (PSF) could be assessed\footnote{This star is not the AO reference object, which is located $\sim$1\,arcmin East.}. Fitting a circular Gaussian PSF core to the image of this star, accounting for the marginal undersampling of the pixels, we estimate a FWHM of 0.2\,arcsec. The fraction of the PSF scattered beyond this core cannot be estimated given the flux of the star.
In the NGC\,5128 field, there are no isolated stars from which the PSF can be directly verified. Instead, we use the collapsed image to compute a two-dimensional spatial autocorrelation function; if only point-sources are present, this process should return the autocorrelation of the PSF. We find the radial profile is consistent with a PSF core of $\sim$0.25\,arcsec FWHM.

\section{Principal Components Decomposition}\label{sec:pca}

Although the collapsed-cube images in Figure~\ref{fig:fields} appear to resolve stars in both galaxies, the signal-to-noise ratio is far too low to extract spectra for individual sources. Instead, we harness the correlated information across the whole datacube, using the methods described in \stt.

We applied the PCA technique to decompose the spectra of the final datacube $S(x,y,\lambda)$ into a linear combination of components ordered by their contribution to the total variance:
\begin{equation}
    S(x,y,\lambda) = \sum_i p_i(x,y) E_i(\lambda) \, ,
\end{equation}
where $E_i(\lambda)$ is the $i$'th eigenspectrum and $p_i(x,y)$ is the $i$'th eigenimage, comprised of the eigenvalues arranged onto the spatial sampling of the datacube.\footnote{Cast in this form, the method is a simple version of 
(unsupervised)
``hyper-spectral unmixing'', a class of methods widely encountered in remote sensing and medical imaging domains \cite[e.g.][]{NIPS1999_798ed7d4,keshava2002spectral,Bioucas}. See \cite{2009MNRAS.395...64S} for a similar approach 
applied to astronomical IFU data.}
The eigenspectra are expected to encode the properties of the brightest stars present in the datacube, which drive the fluctuations, while the corresponding eigenimages show how each components is distributed spatially across the field. (The PCA transformation itself is unaware of the spatial arrangement of the pixels.) 

For application of this method to the ERIS observations, we applied a spatial mask to reject pixels with significantly lower exposure, to ensure similar data quality and provenance across the cube. Likewise, in the wavelength direction, the spectra were restricted to the range 1.49--1.80\,$\mu$m, to avoid steep throughput variation at the edges of the H-band window, and pixels affected by residual noise spikes were masked.

\subsection{NGC\,5128}

\begin{figure}
\includegraphics[width=85mm]{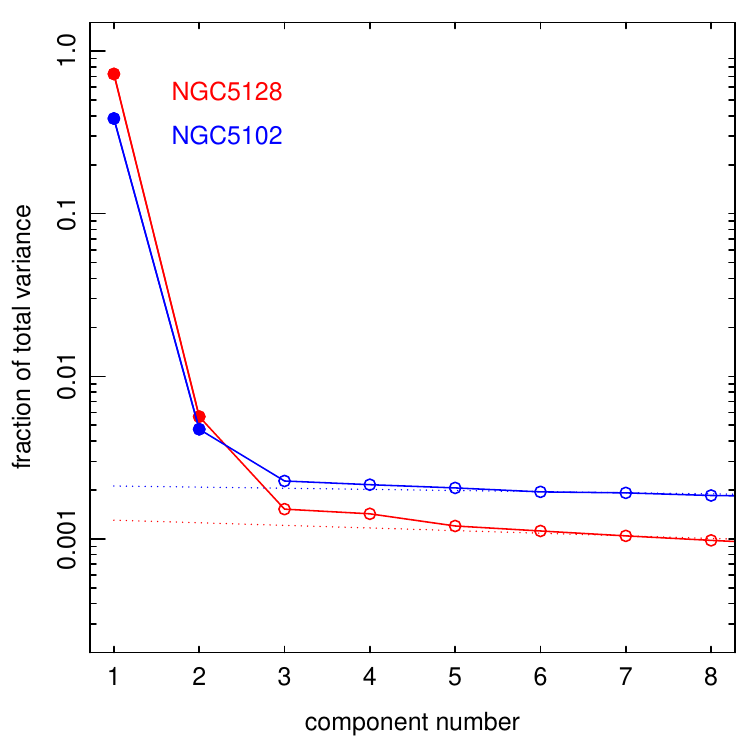}
\vskip -0.5mm
 \caption{Fraction of total variance (including observational noise) contributed by each component in the PCA decomposition for the two galaxies. The dotted lines show fits to components 5--15, representing the noise floor in each observation.
 }
 \label{fig:varcomps}
\end{figure}

For application to NGC\,5128, the final dataset is based on $\sim$4300 spatial pixels with at a minimum 3600\,s exposure (but median $\sim$10000\,s). The principal component variance contributions are shown in Figure~\ref{fig:varcomps}. The information carried by the first two components clearly exceeds the run of contributions from the noise `floor'.

\begin{figure*}
\includegraphics[width=176mm]{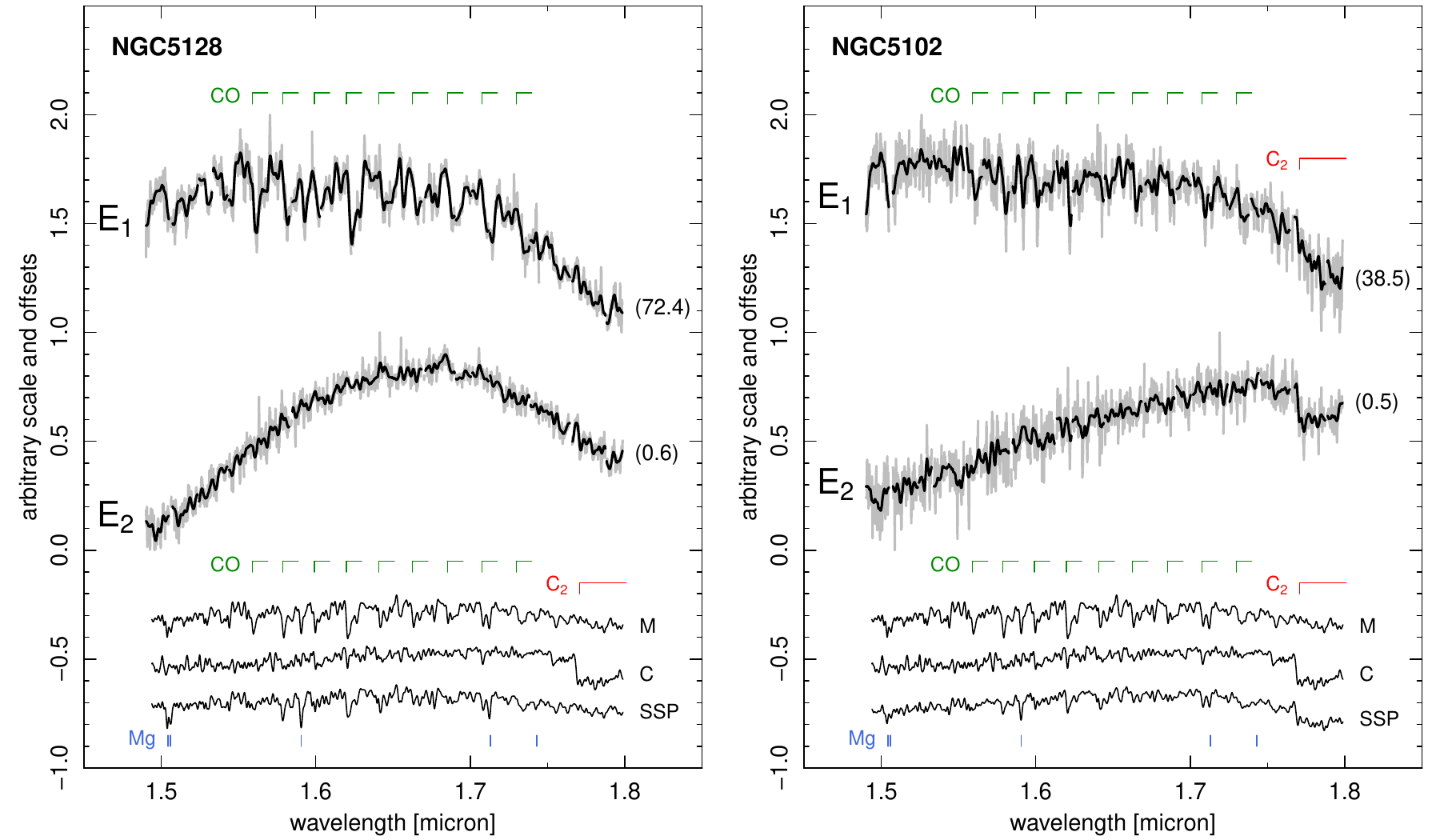}
\vskip -0.5mm
 \caption{The two leading eigenspectra derived for each target field. In parentheses to the right are the percentage contributions to the total data variance accounted for by each component. 
 The spectra are shown at the full resolution of the data (grey), and after smoothing to enhance visibility of features (black).
 The wavelengths of some key molecular bandheads and atomic lines are indicated. Below for comparison, we show composite M-giant and carbon-star spectra, and simple stellar population (SSP) integrated-light spectra from the X-shooter Spectral Library of \protect\cite{2022A&A...661A..50V}. 
 The SSP spectra are for solar metallicity models with  ages of 10\,Gyr (left) and 0.5\,Gyr (right). The comparison spectra have been normalised by a linear continuum.
 }
 \label{fig:eigspec_n5128}
\end{figure*}

The eigenspectra of the two leading components are shown in Figure~\ref{fig:eigspec_n5128}, left. The first component accounts for 72 per cent of the total variance (including observational noise); its eigenspectrum, $E_1$ is dominated by the second-overtone CO bandheads, with the strongest band at $\sim$1.62\,$\mu$m.
The second principal component contributes only 0.56 per cent of the total variance, but this fraction is still notably higher than all subsequent components, which will not be shown or discussed here. The second eigenspectrum $E_2$ does not show any strong narrow features, but instead describes a strongly curved `continuum'. This spectral shape is characteristic of very late giants ($\gtrsim$M7), where it arises from 
the wings of the H$_2$O bands at both sides of the H-band window.

Figure~\ref{fig:eigimg} (upper row) shows the eigenimages corresponding to the first two components. As discussed in S22, in the case of unresolved populations, these images should be structureless, encoding the random sampling of stars in each pixel. In that context, any large scale gradient or structure would be interpreted as a violation of the assumption that all pixels are drawn from the same underlying population. 
As noted already however, in the H-band we are operating in the near- or fully-resolved regime, since the number of very luminous stars per pixel is much smaller than in the optical. The eigenimages demonstrate this clearly, with structure that can be mapped onto the collapsed flux image. Especially for $p_1$, the distribution of eigenvalues is highly non-Gaussian, with a long tail of `bright' pixels driven most likely by individual stars with spectra close to $E_1$, i.e. bright mid-M giants. 

The distribution of the second eigenvalue, $p_2$, is a little more symmetric. As in the MUSE case, the second component should be thought of as a `correction', adding or subtracting $E_2$ from the pixel spectrum. Nonetheless, there are clearly star-like peaks in the second eigenimage which in some cases match up with sources in the flux image but differ from those seen in the first component. We may cautiously interpret these peaks as individual giants with very late M-type spectra. 

All of the results reported here are robust against reasonable changes to the data treatment, e.g. using only pixels with $>$7\,200\,s exposure,
or different criteria for masking the sky lines.

\subsection{NGC\,5102}

For NGC\,5102, we used $\sim$3500 spatial pixels with minimum exposure 1800\,s (median $\sim$3000\,s), having first excluded a small region around the bright star in the field\footnote{In practice the results are unchanged if this star is included.}. As seen in Figure~\ref{fig:varcomps}, there are again two recovered components above the noise level. 

The eigenspectra are shown in Figure~\ref{fig:eigspec_n5128}, right. The first component accounts for 38 per cent of the total variance including the observational noise (which is greater in this case due to the shorter exposure). The corresponding eigenspectrum, $E_1$ again carries the features expected for M-giants, with CO bands noticeable though not so visually prominent as in NGC\,5128.
The second principal component contributes 0.5 per cent of the total variance. Although this is smaller in absolute terms than the equivalent fraction for NGC\,5128, the contribution of the second component relative to the first is larger in NGC\,5102. 
The second eigenspectrum $E_2$ shows a red continuum shape and a sharp break at 1.77\,$\mu$m. This is the head of the \cite{1963ApJ...137...61B} C$_2$ band, which is 
a definitive signature of carbon stars. The weaker break just short-wards, at 1.75\,$\mu$m, is also observed in carbon star spectra, though its molecular or atomic origin is unclear \citep{2017A&A...601A.141G}.

\begin{figure*}
 \includegraphics[width=174mm]{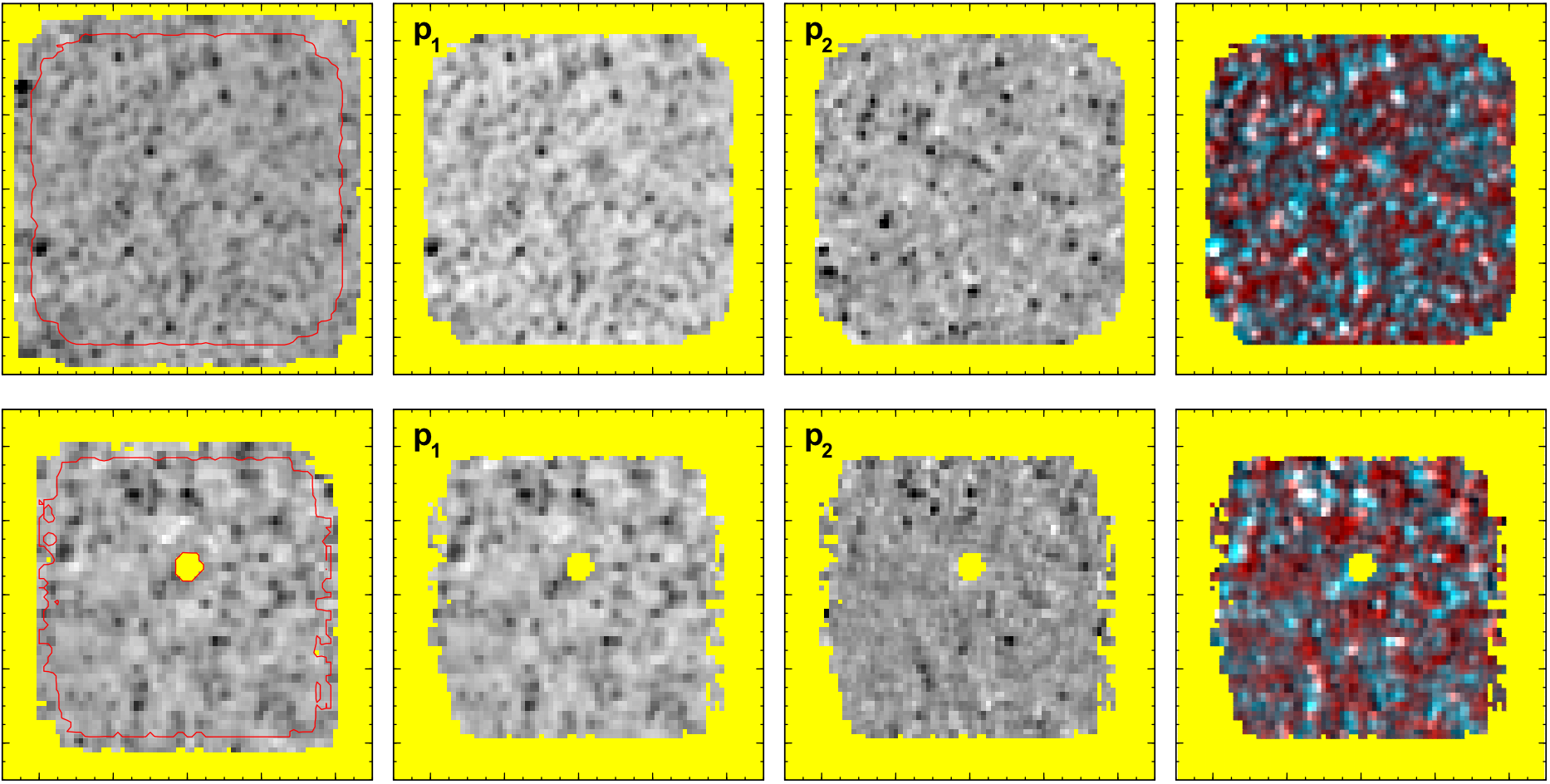}
 \vskip -0.5mm
 \caption{Eigenimages of the PCA decomposition for NGC\,5128 (upper panels) and NGC\,5102 (lower). All panels are 10\,arcsec on a side. For each target, the first panel reproduces the collapsed-cube image for comparison  (with the bright star excised in the case of NGC\,5102), with a contour showing the pixels selected for the PCA. The second and third panels show the first two eigenimages in greyscale form, while the final panel shows a two-colour visualisation, with $p_1$ mapped to the blue channel and $p_2$ to red. The resulting image effectively maximizes the information carried; no selection of broad or narrow filters could discriminate the stellar content so clearly.}
 \label{fig:eigimg}
\end{figure*}

The eigenimages for the two components for NGC\,5102 are shown Figure~\ref{fig:eigimg} (lower row). Again, the images confirm that components are distributed in compact sources, with no evidence of large-scale spatial gradients or uncorrected instrumental signatures.

\section{Discussion}\label{sec:disc}

In both NGC\,5128 and NGC\,5102, the PCA decomposition has recovered two distinct components present among the bright stellar content. For both galaxies, the first component is that expected from models, being broadly the spectrum of mid-M stars on the giant branch. 

The second component in NGC\,5128 was unexpected prior to the analysis. Our Poisson-sampled simple stellar population models, as presented in GS26, do not yield such a feature (see Appendix~\ref{sec:mock}), and nor do composite populations including age or metallicity spreads. As noted above, the spectral shape of $E_2$ is characteristic of very cool giant stars, so the implication is that a population of such stars must be  present in NGC\,5128 at a greater level than predicted. One possibility is that cooler phases in the pulsation cycles of Mira-like long-period variable (LPV) stars are responsible for this component. As shown, for example, in \cite{2000A&AS..146..217L}, the H-band spectra of such stars vary strongly with phase, with the H$_2$O bands strengthening near optical minimum-light, causing a strongly convex `continuum' shape, while the CO bandheads weaken. The  typical flux variation of LPVs is much smaller in the near-IR than in the optical (0.5--1.0\,mag in H), so the contribution of the `faint-phase' spectrum to the total light may not be negligible.

In general many cool AGB stars are expected to be LPVs, an effect which will be averaged over the population for integrated light spectroscopy, but which will introduce stochasticity over the small luminosity sampled by each spatial pixel here.  \cite{2003A&A...411..351R} found that most of the bright stars above the red giant branch tip in NGC\,5128 were variables, but most of these will belong to the intermediate-age stellar population, and be rare in the small field observed here. We suggest that the $E_2$ signal instead arises from fainter LPVs in the bulk old bulge population. 
Appendix~\ref{sec:muselpv} presents supporting evidence for many faint candidate LPVs in  NGC\,5128, from the \stt\ MUSE observations.
As an example in another old, metal-rich environment, \cite{2015Natur.527..488C} found pixel-level variability in multi-epoch HST observations of M87, which could be reproduced with models including LPVs.
Ultimately, our hypothesis that variable stars drive $E_2$ in NGC\,5128 is testable: future second-epoch observations should recover the same eigenspectra and ensemble behaviour, but sample the variables at different phases, with different pixels lighting up in the eigenimages at different epochs. 
The overall incidence of the cool-phase LPVs can only be quantified by including them directly into our Poisson-sampled models; doing so would also allow us to estimate their contribution to the integrated light spectra of old galaxies. This will be undertaken in a future work.

By contrast, the detection of the second fluctuation component in NGC\,5102 was foreseen, and was one of the goals of the observation.  The optical spectrum of this S0 galaxy shows strong post-starburst signatures (blue continuum, deep Balmer absorption lines), especially in the nucleus, while the rest of the galaxy has a characteristic age of $\sim$2\,Gyr, which likely conceals a spread extending below 1\,Gyr  \citep{2011ApJ...727L..15M,2015ApJ...799...97D,2017MNRAS.464.4789M}.
At these ages, the population will include AGB stars with initial masses $\gtrsim$1.5\,$M_{\odot}$, which undergo thermal-pulse-driven dredge-up, converting some of them to carbon stars \citep[e.g.][]{1995A&A...293..381G}. Many works have demonstrated the presence of carbon stars in $\sim$1\,Gyr integrated-light spectra ranging from clusters in the Large Magellanic Cloud \citep[e.g.][]{2012A&A...543A..75L} out to massive galaxies at $z$\,=\,1--2 \citep[e.g.][]{2025NatAs...9..128L}. 

The carbon star features of post-starburst galaxies have been used in efforts to constrain
the contribution of the TP-AGB phase to $\sim$1\,Gyr populations, which is a notably uncertain prediction of spectral synthesis models. \cite{2013MNRAS.428.1479Z} reported no detection of the 1.77\,$\mu$m C$_2$ band, in contrast to predictions from TP-AGB-heavy models such as those of \cite{2005MNRAS.362..799M}.
Likewise, the carbon-specific features were weak, and uncorrelated with age, in  the spiral galaxies studied by \cite{2015MNRAS.450.3069R}, though other features possibly attributable to TP-AGB stars were detected.
Subsequent studies \citep[e.g.][]{Baldwin2018,DahmerHahn2018} highlighted the difficulty of finding consistent fits to the optical and near-IR integrated-light spectra using current models, and hence alternative perspectives such as the method presented here  should provide useful additional constraints.

In NGC\,5102 itself, \cite{2010AJ....139..680D} detected bright carbon stars photometrically in the near-IR, while 
\cite{2011ApJ...727L..15M} measured a spectral break feature at 1.77\,$\mu$m from the integrated light, to infer their presence. As shown, however in the lower part of Figure~\ref{fig:eigspec_n5128}, the effect of these stars in the model spectrum is fairly subtle even for wholly-young populations. 
The dramatic difference in spectrum between O-rich and C-rich atmospheres among the bright giant stars is well suited to detection with our PCA approach. Hence the recovered eigenspectrum $E_2$ cleanly isolates the carbon star features from the datacube\footnote{This is not the same as isolating `the spectrum of carbon stars': $E_2$  functions as a correction term, adding to the features already described by $E_1$.}.
The models presented in GS26 do not currently extend to the young age ranges needed to address the incidence of carbon stars quantitatively. Again, future work will investigate this; by augmenting the isochrones with empirical TP-AGB spectra, it should be possible to establish the characteristic luminosity of the stars which carry the C$_2$ features.

We close by emphasizing some differences between the results shown here and those of \stt, in light of future application to more distant galaxies. 
The optical (approximately I-band) observations of \stt\ sampled the population in the high-$\bar{N}$ regime, where $\bar{N}$ is the effective number of stars per spectrum,  relative to the SBF flux \citep{1993A&A...275..433B}. In the I-band, the 0.125\,arcsec binned pixels used in \stt\ yield $\bar{N}$\,$\approx$\,30. Thus although Poisson variation imprints the signal of the giant stars in the data, the eigenimages appeared featureless. By contrast in the ERIS observations here, $\bar{N}$\,$\approx$\,3--5 in the same pixel size, and the maps of $p_1$ and $p_2$  reveal individual stars above the fluctuating background.
This is a reflection of the intrinsic strength of the fluctuations being greater in the H-band, just as photometric SBFs are more easily measured in the IR. 
Likewise, it is partly because the fluctuations are stronger that we are able to recover two distinct components of variation where only one was detected with MUSE, though the intrinsic stellar mixture also contributes.

Looking ahead to 
future instruments such as HARMONI on the Extremely Large Telescope \citep{Thatte2022}, at least a factor-of-five resolution gain is possible, such that similar physical sampling can be achieved in Virgo or Fornax as reached here for Centaurus, at similar surface brightness.
We note however, however, that extracting the fluctuation components does not  require the galaxy to be near-resolved. Just as for the classical SBF, our method is just as applicable in the high-$\bar{N}$ bright cores of true elliptical galaxies, which will never be strictly resolvable.
All that is needed is to detect sufficient photons per individual star. In this scenario, as well as for the low-$\bar{N}$ case, the 30-m class telescopes will be needed to reach Virgo distances. The closest true elliptical, NGC\,3379, should however be within reach with deep integrations using ERIS.

\section{Conclusions}\label{sec:concs}

We have extended the method of PCA decomposition applied to IFU spectroscopy into the near-infrared, analysing observations for two contrasting galaxies in the Centaurus Group.  Our main conclusions are:

\begin{itemize}

\item Our ERIS H-band data are able to resolve the brightest stars in the collapsed images, but without sufficient signal-to-noise to obtain spectra directly for individual sources. 

\item
In each galaxy, the pixel-to-pixel differences in spectrum can be described by two principal components, with different spectral shapes. For both targets, the leading component has, as expected, a typical M-type spectrum as previously found in the red-optical (\stt) and as predicted by models (GS26). 

\item
In both galaxies, the PCA method recovers a second component, contributing a much smaller fraction to the total variance in the data. The second eigenspectrum is physically interpretable in each case. The ability to detect such second (or further) components is a unique advantage of our more general analysis method, as compared to the approaches of \cite{2014ApJ...797...56V} or \cite{2018MNRAS.480..629M}.

\item
In the old bulge of NGC\,5128, the second component describes contributions from very cool giants, which arise from Mira-like variables in the faint phases of the pulsation cycles. Future second-epoch observations would be able to test this conjecture.

\item
In the post-starburst galaxy NGC\,5102, the second component instead carries features attributable to carbon stars among the AGB population. Such stars were known to be present from integrated light studies \cite{2011ApJ...727L..15M} but are here isolated much more cleanly through the fluctuation eigenspectrum approach.

\end{itemize}

In summary, we have shown that harnessing the pixel-to-pixel variance between spectra in H-band IFU data can reveal qualitatively new information about the stellar content of galaxies, even when individual stars cannot be traditionally resolved. A more quantitative interpretation of our results, e.g. constraining the number and luminosity of the faint-phase LPVs in NGC\,5128 or carbon stars in NGC\,5102, requires extensions to our Poisson-sampled population models, which will be a focus for ongoing work. Our results confirm the feasibility of applying this technique to more distant galaxies, with the closest true elliptical galaxy, NGC\,3379 at 10\,Mpc, being the most enticing target.

\section*{Acknowledgements}

This work is based on observations collected at the European Organisation for Astronomical Research in the Southern Hemisphere under ESO programme 115.28DN. Both authors were supported by the STFC through consolidated grant ST/X001075/1. We are grateful to Marina Rejkuba for initially prompting the analysis presented in  Appendix~\ref{sec:muselpv}.

\section*{Data availability}

The primary data used in this paper are available from the ESO public archive, with programme number 115.28DN. The supporting datasets used are also publicly available from the 2MASS, ESO and HST archives. The Poisson-sampled stellar population models used for the analysis in Appendix~\ref{sec:mock} will be published in GS26.

\bibliographystyle{mnras}
\bibliography{bibl} 

\appendix

\section{NGC\,5128 model comparison}\label{sec:mock}

\begin{figure*}
\includegraphics[width=177mm]{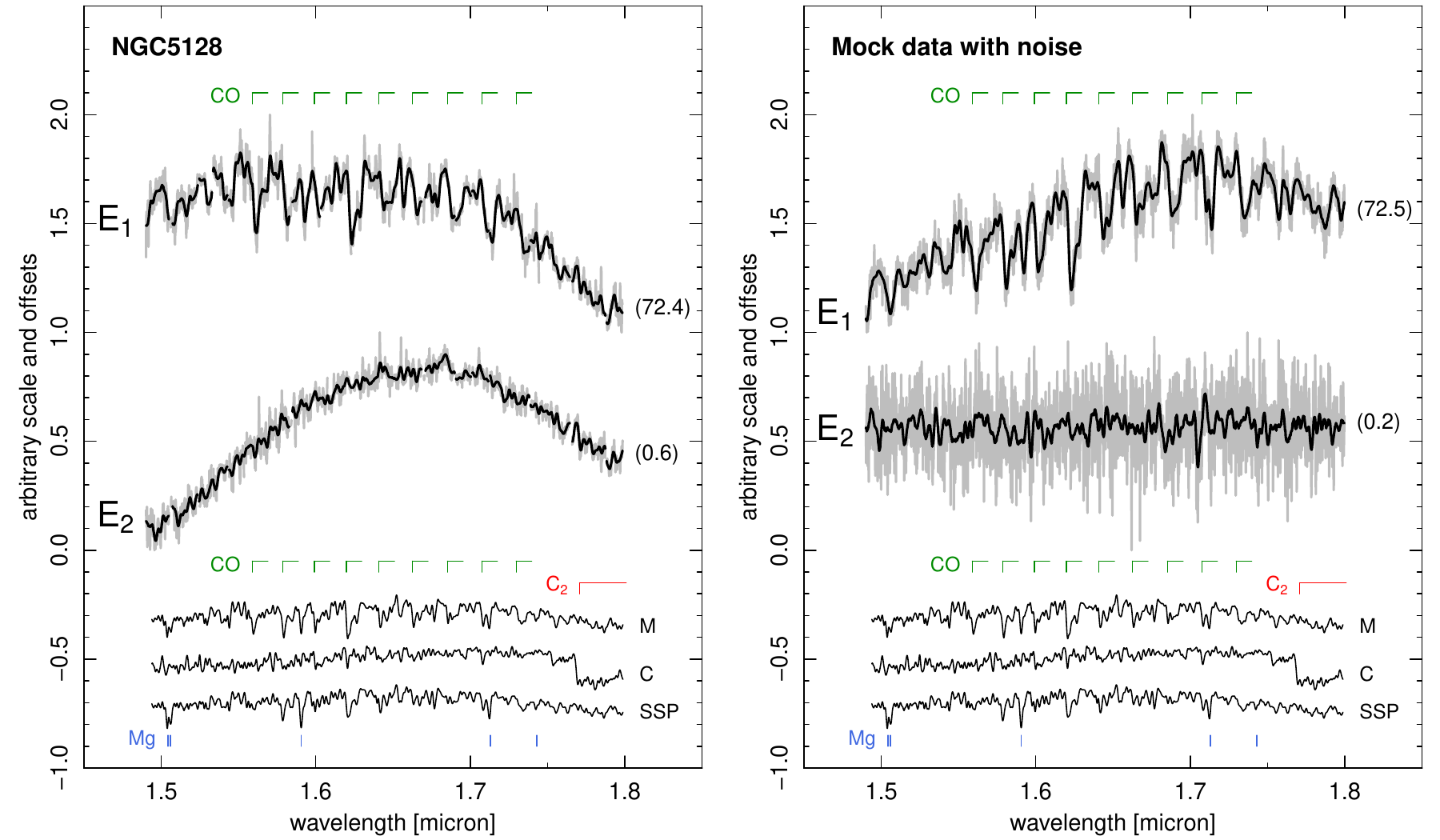}
\vskip -1mm
 \caption{Comparison of eigenspectra for the observed NGC\,5128 data (left, reproduced from Figure~\ref{fig:eigspec_n5128}, for ease of reference) and for a mock dataset derived from Poisson-sampled stellar population models from GS26 (right). 
 }
 \label{fig:eigspec_mock}
\end{figure*}

\begin{figure}
\includegraphics[width=83mm]{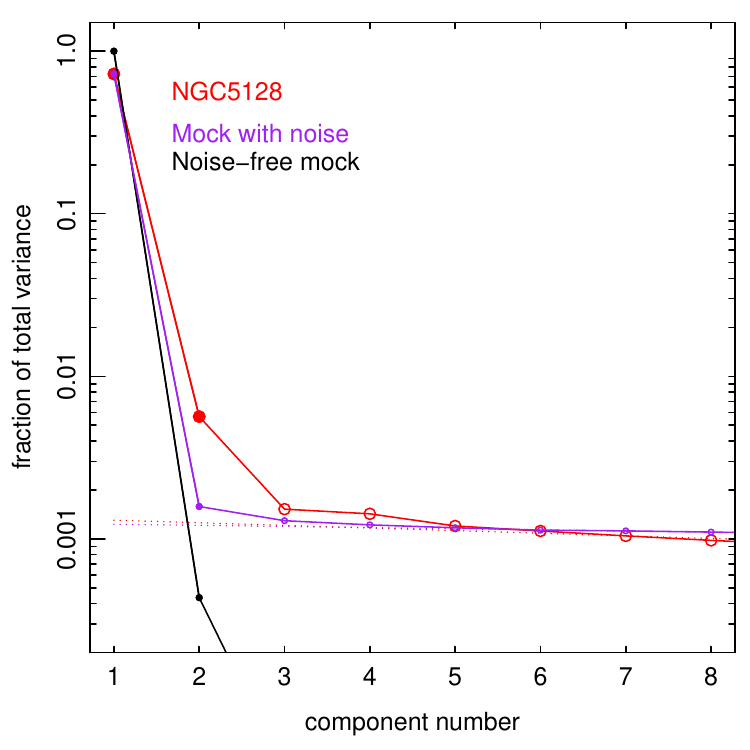}
\vskip -1mm
 \caption{The PCA variance fractions for  NGC\,5128, compared to mock data generated from the Poisson-sampled spectral synthesis models of GS26. The purple line corresponds to the same mock dataset used for Figure~\ref{fig:eigspec_mock}, while the black shows the equivalent if no noise is added to the mocks.}
 \label{fig:varcompsmock}
\end{figure}

\begin{figure*}
\includegraphics[width=178mm]{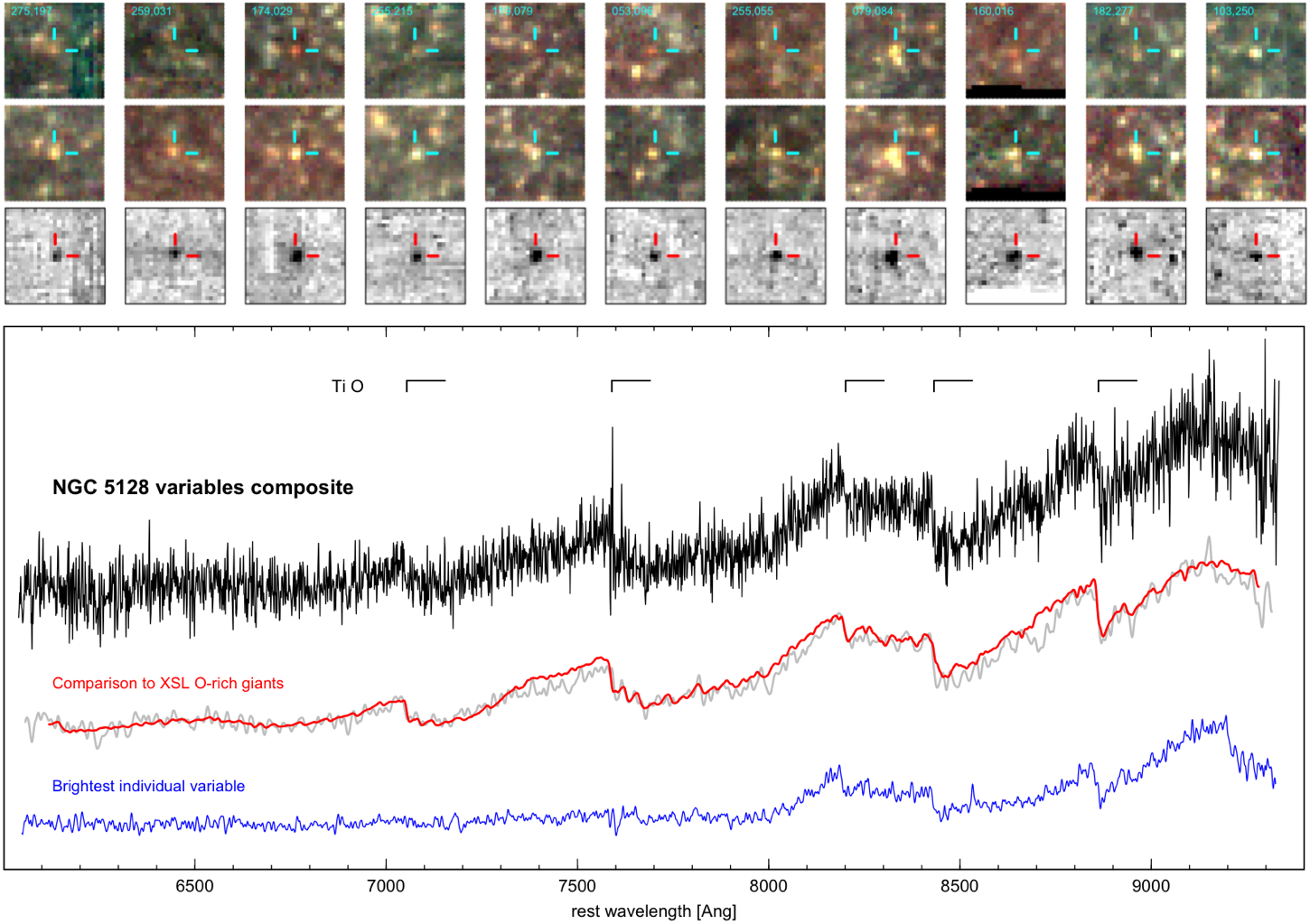}
\vskip -0mm
 \caption{ Upper section: cut-out images of 11 candidate long-period variable stars selected from the  \stt\ two-epoch MUSE Narrow Field Mode observations. The first row shows a colour rendering of the epoch in which each star is fainter, and the second shows the epoch in which it is brighter. Each panel shows a region 0.625\,arcsec on a side.
The colour channels are blue = 7475--8100\,\AA; green = 8100--8725\,\AA; red = 8725--9350\,\AA. The third row shows a greyscale difference image over the combined range, 7475--9350\,\AA. Lower section: composite spectrum of the 11 LPV candidates,
and a comparison to the best-matching composite of O-rich cool giants from 
\citep{2022A&A...661A..50V}. 
For visual comparison, both the XSL spectrum (red) and the NGC\,5128 variables composite (grey) have been smoothed by $\sigma$\,=\,150\,km\,s$^{-1}$. Below we show the smoothed spectrum of the brightest individual LPV candidate  
(eighth from left in upper panels). 
}
 \label{fig:varbs}
 \end{figure*}

As noted in the main text, the second fluctuation  component detected in our ERIS data for NGC\,5128 is not predicted by the  Poisson-sampled spectral synthesis models. The models will be described in detail by GS26, but are essentially an extension of the method of \citep{2022MNRAS.509.5737S} to cover a broader range both in wavelength and in stellar population age and metallicity.

To demonstrate the use of these models for comparison to data, we generated 1000 Poisson-sampled realisations of the H-band spectrum of a 10\,Gyr, solar-metallicity population, with a surface brightness matched to the NGC\,5128 observations. Each spectrum was shifted and smoothed to approximately the appropriate redshift and velocity dispersion for this galaxy, resampled onto the ERIS wavelength bins, and a uniform noise was added to approximate the observations. 

Analysing these mock spectra with exactly the same PCA process used for the real data, we recover the eigenspectra shown in Figure~\ref{fig:eigspec_mock}. In the first component, although the `continuum' shape 
differs, the model reproduces the narrow features fairly closely. (No attempt was made to tune the model parameters to improve the match.) In contrast, the second eigenspectrum for the mock data shows no narrow spectral features, nor any smooth component with a different shape from $E_1$, which is the salient characteristic of the observed data. 
The variance contributed by the second component is a factor of 3.5 smaller in the mock data than the equivalent for the observations, and essentially consistent with the noise `floor' (Figure~\ref{fig:varcompsmock}). In the absence of noise, the models predict the second component to contribute an order of magnitude less variance than the leading term, and hence would be below detectability with our observations.

\section{Long-period variables in NGC\,5128}\label{sec:muselpv}

In Section~\ref{sec:disc} we suggested that the second component in NGC\,5128 might be related to the cooler phases of long-period variable stars. In this appendix, we take the opportunity to present related and previously unpublished results from the MUSE Narrow Field Mode observations of \stt.

By chance, the \stt\ data were acquired across two observing seasons, with data taken 
on 2020 Feb 4, 2020 Mar 17, and 2021 Apr 9 (two exposures). Combining the 2020 observations as a single first epoch, and the 2021 observations as a second epoch, gives an effective baseline of 409 days.
The two datasets have similar angular resolution, with an effective core FWHM of 0.063\,arcsec in the first epoch and 0.055\,arcsec in the second, estimated by fitting a three-gaussian model to the on-axis AO reference star. The AO star was also used to provide an approximate photometric calibration for each MUSE cube, based on its {\it Gaia} DR3 measurements, transformed to the I band.

The \stt\ MUSE reconstructed images are subject to various image artefacts and imperfect astrometric registration across the field of view. Given these effects, and the high degree of crowding, a strictly rigorous selection of variables
was not attempted. Instead,
variable star candidates were identified through visual comparison of I-band images for the two epochs. 
The process used a combination of blinking and difference-image methods, using only the image area covered by both epochs to the full depth. Eleven of the most convincing variable candidates are shown in the upper part of Figure~\ref{fig:varbs}.

We extracted spectra from the brighter epoch of each LPV candidate, using a small aperture (radius 0.075\,arcsec) to minimise contamination from neighbouring stars, and subtracting a local background from a  0.2--0.3\,arcsec annulus.
Of the 11 stars, only
 one  
individually yields a meaningful spectrum,  showing prominent Ti\,O bands characteristic of an O-rich giant, and no detectable flux at wavelengths below 8000\,\AA\ (lower part of Figure~\ref{fig:varbs}). 
Cross correlation against library templates yields a velocity of 320\,km\,s\,$^{-1}$, consistent with membership of NGC\,5128.
This star has $I$\,$\approx$\,22 in the brighter epoch, some two magnitudes above the tip of the old first-ascent giant branch. Hence it is likely an AGB star drawn from the younger sub-population in this galaxy \citep{2001A&A...379..781R}.
The remaining LPV candidates have $I$\,$\gtrsim$\,23, and their spectra are much noisier. 
We therefore created a composite spectrum from all 11 stars, combined using equal weighting so that the brightest one does not dominate the stack. The average spectrum is a  close match to 
a $I-K$\,=\,4.23 O-rich giant composite from the X-Shooter Spectral Library 
\citep[XSL;][]{2022A&A...661A..50V}. (The strongest deviation, at $\sim$8710\,\AA, does not seem to correspond to a known feature in cool stars.)

\bsp	
\label{lastpage}
\end{document}